\documentstyle[aps]{revtex}

\begin{document}
\title{Main Parameters of the Linac-Ring Type Phi, Charm and Tau Factories }
\author{A.K. \c{C}ift\c{c}i$^{a}$, E. Recepo\u{g}lu$^{a},$ S. Sultansoy$^{b,c},$
\"{O}. Yava\c{s}$^{d}$}
\address{$^{a}$Dept. of Physics, Faculty of Sciences, Ankara University, 06100\\
Tandogan, Ankara, TURKEY\\
$^{b}$Physics Dept., Faculty of Arts and Sciences, Gazi University, 06500\\
Teknikokullar, Ankara, TURKEY \\
$^{c}$Institute of Physics, Academy of Sciences, H. Cavid Ave. 33, Baku,\\
AZERBAIJAN\\
$^{d}$Dept. of Eng. Physics, Faculty of \ Engineering, Ankara University,\\
06100 \\
Tandogan, Ankara, TURKEY}
\maketitle

\begin{abstract}
Parameters for linac-ring type $e^{-}e^{+}$ colliders with $\surd s=1020$
MeV ($\phi $ factory), 3770 MeV (charm factory) and 4.2 GeV ($\tau $
factory) are discussed. It is shown that luminosities of the order of 10$%
^{34}$ cm$^{-2}$s$^{-1}$ can be achieved.
\end{abstract}

\section{Introduction}

An old idea of colliding of the electron beam from a linac with a beam
stored in a ring [1] is widely discussed during the last decade with two
purposes:

1) to achieve the TeV energy scale in lepton-hadron and photon-hadron
collisions (see review articles [2] and references therein),

2) to construct high luminosity particle factories, namely, B-factory [3], $%
\phi $-factory [4,5], $c-\tau $-factory [6] etc.

\bigskip

Concerning the first direction, TESLA$\otimes $HERA based $ep$, $\gamma p$, $%
eA$ and $\gamma A$ colliders are included in TESLA project [7]. And, Linac$%
\otimes $LHC based $ep$, $\gamma p$, $eA,$ $\gamma A$ and $FEL\gamma A$
colliders [8] can be considered as the next step. On the other hand,
linac-ring type B-factory lose its attractiveness with coming into operation
of KEK-B [9] and PEP-B [10] colliders.

\bigskip

In this paper, we show that linac-ring type particle factories still is the
matter of interest taking in mind $\phi ,$ charm and $\tau $ options. In
section 2, we present a general consideration of beam dynamics aspects of
linac-ring type colliders. Proposed parameters for linac-ring type $\phi $
factory are given in section 3A where we also compare it with DA$\Phi $NE
[11]. Linac-ring type charm and $\tau $ factories are discussed in sections
3B and 3C respectively. In the final section, we give some concluding
remarks.

\bigskip

\section{General Considerations}

From the point of view of particle physics there are two most important
collider parameters: center of mass energy and luminosity. For
ultra-relativistic colliding beams, center of mass energy is given by

\begin{equation}
\sqrt{s}=2\sqrt{E_{1}E_{2}}
\end{equation}
In our case, E$_{1}$ is the energy of electrons accelerated in linac and E$%
_{2}$ is the energy of positrons stored in ring. For $\phi $ and charm
factories, it is important to have $\Delta (\surd s)<\Gamma $ in order to
use the advantage of resonant production of $\phi $ and $\psi (3S)$ mesons: $%
m_{\phi }=1019.417\pm 0.014$ MeV \ with $\Gamma _{\phi }=4.458\pm 0.032$ MeV
and $m_{\psi (3S)}=3769.9\pm 2.5$ MeV with $\Gamma _{\psi (3S)}=23.6\pm 2.7$
MeV [12]. This condition is not so crucial for $\tau $ factory because of
pair production of $\tau $ leptons.

The luminosity of $e^{-}e^{+}$ collisions is given by

\begin{equation}
L=\frac{N_{e}N_{p}}{2\pi \sqrt{(\sigma _{xe}^{2}+\sigma _{xp}^{2})(\sigma
_{ye}^{2}+\sigma _{yp}^{2})}}f_{c}H_{D}
\end{equation}
where $N_{e}$ is number of electrons per bunch, $N_{p}$ is number of
positrons per bunch, $\sigma _{x,y}$ are horizontal and vertical beam sizes, 
$f_{c}$ is collision frequency. H$_{D\text{ }}$is luminosity enhancement
factor which is calculated by using GUINEA-PIG beam- beam simulation program
[13] .

The first restrictive limitation for electron beam is beam power

\begin{equation}
P_{e}=N_{e}E_{e}f_{c}
\end{equation}
which determines the maximum value of $N_{e}f_{c}$ in Eq. (2).

The maximum number of electrons per bunch is determined by the beam-beam
tune shift limit of the positron beam

\bigskip 
\begin{equation}
\Delta Q_{p}=\frac{N_{e}r_{0}}{2\pi \gamma _{p}}\frac{\beta _{p}^{\ast }}{%
\sigma _{xe}(\sigma _{xe}+\sigma _{ye})}
\end{equation}
where $r_{0}=2.81\times 10^{-15}$ m is the classical radius of the electron, 
$\gamma _{p}$ is the Lorentz factor of the positron beam and $\beta
_{p}^{\ast }$ is beta function at collision point. Generally accepted
beam-beam tune shift value for positrons in case of ring-ring colliders is $%
\Delta Q\leq 0.06.$ This limit value can be a little bit larger for
linac-ring type colliders.

Parameters of positron beams are constrained by the disruption D of
electrons which is\ defined as the ratio of the positron bunch length to the
electron focal length

\begin{equation}
D_{ye}=\frac{\sigma _{zp}}{f_{ye}}=\frac{2r_{0}N_{p}\sigma _{zp}}{\gamma
_{e}\sigma _{yp}(\sigma _{xp}+\sigma _{yp})}
\end{equation}
where $\gamma _{e}$ is the Lorentz factor of the electron beam and $\sigma
_{zp}$ is the positron bunch length. In this study we consider the round
beam case: $\sigma _{ze}=\sigma _{ye}=\sigma _{e}$ and $\sigma _{xp}=\sigma
_{yp}=\sigma _{p}.$ The analysis performed for linear colliders shows that $%
D_{xe}=D_{ye}=D=25$ is acceptable [14].

\section{Parameter Sets}

\subsection{Linac-Ring Type $\protect\phi $ Factory}

In order to make the linac-ring type $\phi $ factory feasible, its
luminosity should exceed the luminosity of DA$\Phi $NE (operating standard
ring-ring type $\phi $ factory at Frascati) at least by one order. The
design luminosity of DA$\Phi $NE is 5.3$\times 10^{32}$ cm$^{-2}s^{-1}$,
however only $L=$1.8$\times 10^{31}$cm$^{-2}s^{-1}$ has been achieved since
start in 1999 [11]. Therefore, $L=$5$\times 10^{33}$ cm$^{-2}s^{-1}$ for
linac-ring type machine will be quite enough, even if the design luminosity
value would be achieved in DA$\Phi $NE. In Table I, we present proposed
parameters for a number of linac-ring type $\phi $ factory options.
Concerning the energy of e$^{-}$ and e$^{+}$ beams, we consider two options E%
$_{e^{-}}=130$ MeV and 260 MeV (according to Eq. (1), E$_{e^{+}}=2$ \ GeV
and 1 GeV, respectively).

\bigskip As mentioned in section II, in the case of $\phi $ and charm
factories, it is important to obey condition $\Delta (\surd s)<\Gamma .$ The
expected luminosity spectrum $dL/dW_{cm}$ for option C in Table I is plotted
in Figure 1. We have used GUINEA-PIG simulation program [13] with $\Delta
E_{e^{+}}/E_{e^{+}}=\Delta E_{e^{-}}/E_{e^{-}}=10^{-3}.$ One can see that
center of mass energy spread is well below $\Gamma _{\phi }\approx 4.46$
MeV. Therefore, we can use the well known Breit-Wigner formula 
\begin{equation}
\sigma _{BW}=\frac{12\pi }{m_{\phi }^{2}}B_{in}B_{out}
\end{equation}
where $B_{in}$ and $B_{out}$ are the branching fractions of the resonance
into the entrance and exit channels. Branching fractions for different decay
modes of the $\phi $ meson [12] are given in Table II. According to Eq. (6)
we expect about 4$\cdot 10^{11\text{ }}$events per working year (10$^{7}$
s). Number of events for different decay modes are presented in the last
column of Table II.

\bigskip

\subsection{Linac-Ring Type Charm Factory}

\bigskip Recently CLEO-c (see [15] and references therein) proposal has been
approved in order to explore the charm sector starting early 2003. With a
necessary upgrade, expected machine performance for CLEO-c will be 3$\times
10^{32}$cm$^{-2}s^{-1}$ at $\surd s=3.77$ GeV. In Table III, we present
proposed parameters for three different linac-ring type charm factory
options. As one can see, $L=10^{34}$cm$^{-2}s^{-1}$ can be achieved which
exceeds the CLEO-c design luminosity by more than an order.

In Figure 2, we plot expected luminosity spectrum for option C from Table
III. It is seen that center of mass energy spread is well below $\Gamma
_{\psi (3S)}\approx 24$ MeV. Using the Eq. (6) with replacement of $m_{\phi
} $ to $m_{\psi (3S)}$ and $Br(\psi (3S\rightarrow e^{+}e^{-})\approx
10^{-4},$ we obtain the expected number of $\psi (3S)$ per working year,
which is about 10$^{10}.$ Let us mentioned that $D\overline{D}$ mode is the
dominant one for $\psi (3S)$ decays. An additional advantage of the proposed
charm factory is the asymmetric kinematics. This feature will be important
in investigations of $D^{0}\overline{D^{0}}$ oscillations and CP-violations
in charmed particle decays.

\subsection{Linac-Ring Type $\protect\tau $ Factory}

The cross section of the process $e^{+}e^{-}\rightarrow \tau ^{+}\tau ^{-}$
for $s<<m_{\tau }^{2}$ is given by

\begin{equation}
\sigma =\frac{2\pi }{3}\frac{\alpha ^{2}}{s}\beta (3-\beta ^{2})\approx 
\frac{43.4\text{ nb}}{s\text{ (GeV}^{2}\text{)}}\beta (3-\beta ^{2})
\end{equation}
where $\beta =\sqrt{1-4m_{\tau }^{2}/s}$ and $\alpha $ is the fine structure
constant. The maximum value of $\sigma =3.56$ nb is achieved at $\surd
s\approx 4.2$ GeV. In difference from $\phi $ and charm factories, in the
case of $\tau $-factory we have consider the symmetric option (E$_{e^{-}}=$E$%
_{e^{+}}=2.1$ GeV). Proposed set of parameters is given in Table IV. One can
see that linac-ring type $\tau $-factory will produce \ $\sim 4.6\cdot
10^{8} $ $\tau ^{+}\tau ^{-}$ pair per working year, which exceeds by two
order the statistics obtained at LEP and CLEO up to now.

\section{Conclusion}

We have shown that linac-ring type machines will give an opportunity to
achieve $L=10^{34}$ cm$^{-2}$s$^{-1}$, which essentially exceeds the
luminosity values of existing and proposed standard (ring-ring type) $\phi $%
, charm and $\tau $ factories. This leads to an obvious advantage in search
for rare decays. Another important feature of linac-ring type $\phi $ and
charm factories is the asymmetric kinematics. This will be important in
investigation of oscillations and CP-violation in strange and charm sector
of the SM.

\bigskip

This work is partially supported by Turkish State Planing Organization (DPT).

\bigskip

\bigskip 
\begin{table}[tbp] \centering%
%
\caption{Main parameters of $\phi $ factory\label{key}}

\begin{tabular}{llll}
& Option A & Option B & Option C \\ \hline
E$_{e}$ (GeV) & 0.130 & 0.130 & 0.260 \\ 
E$_{p}$ (GeV) & 2 & 2 & 1 \\ 
N$_{e}$ (10$^{10}$) & 0.6 & 1 & 0.6 \\ 
N$_{p}$ (10$^{10}$) & 20 & 20 & 10 \\ 
f$_{c}$ \ \ \ (MHz) & 30 & 30 & 30 \\ 
$\beta _{e}/\beta _{p}$ \ \ \ \ (cm) & 0.25/0.25 & 0.25/0.25 & 0.25/0.25 \\ 
$\sigma _{e}/\sigma _{p}$ \ \ \ \ ($\mu m$) & 3.77/9.44 & 4.95/9.44 & 
5.33/4.65 \\ 
$\varepsilon _{e}^{N}/\varepsilon _{p}^{N}$ \ \ \ \ ($\mu mraad$)\  & 
1.45/140 & 2.5/140 & 5.8/17 \\ 
$\sigma _{ze}/\sigma _{zp}$ \ \ (cm) & 0.1/0.1 & 0.1/0.1 & 0.1/0.1 \\ 
$\Delta Q$ & 0.06 & 0.06 & 0.06 \\ 
D & 24.77 & 24.77 & 25.5 \\ 
H$_{D}$ & 1.05 & 1.17 & 1.8 \\ 
L \ \ \ (10$^{34}$ cm$^{-2}s^{-1}$) & 0.6 & 1 & 1.1 \\ 
Linac beam power \ (MW) & 3.74 & 6.24 & 7.48
\end{tabular}
\end{table}%
%

\bigskip 
\begin{table}[tbp] \centering%
%
\caption{ Rates of $\phi $ decays at linac-ring type $\phi $
factory\label{key}}

\begin{tabular}{lll}
Decay modes & Branching ratios & N$_{ev}/year$ \\ \hline
$K^{+}K^{-}$ & 0.492 & 1.968$\cdot 10^{11}$ \\ 
$K_{L}^{0}K_{S}^{0}$ & 0.338 & 1.352$\cdot 10^{11}$ \\ 
$\rho \pi $ & 0.155 & 0.620$\cdot 10^{11}$ \\ 
$\eta \gamma $ & 1.3$\cdot 10^{-2}$ & 5.200$\cdot 10^{9}$ \\ 
$\pi ^{0}\gamma $ & 1.26$\cdot 10^{-3}$ & 5.040$\cdot 10^{8}$ \\ 
$e^{+}e^{-}$ & 2.91$\cdot 10^{-4}$ & 11.64$\cdot 10^{7}$ \\ 
$\mu ^{+}\mu ^{-}$ & 3.7$\cdot 10^{-4}$ & 14.80$\cdot 10^{7}$ \\ 
$\eta e^{+}e^{-}$ & 1.3$\cdot 10^{-4}$ & 5.200$\cdot 10^{7}$ \\ 
$\pi ^{+}\pi ^{-}$ & 7.5$\cdot 10^{-5}$ & 30$\cdot 10^{6}$ \\ 
$\eta (958)\gamma $ & 6.7$\cdot 10^{-5}$ & 26.8$\cdot 10^{6}$ \\ 
$\mu ^{+}\mu ^{-}\gamma $ & 1.4$\cdot 10^{-5}$ & 5.6$\cdot 10^{6}$%
\end{tabular}
\end{table}%
%

\bigskip 
\begin{table}[tbp] \centering%
%
\caption{ Main parameters of charm factory\label{key}}

\begin{tabular}{llll}
\  & Option A & Option B & Option C \\ \hline
E$_{e}$ (GeV) & 1 & 1.18 & 1.42 \\ 
E$_{p}$ (GeV) & 3.55 & 3 & 2.5 \\ 
N$_{e}$ (10$^{10}$) & 0.1 & 0.1 & 0.1 \\ 
N$_{p}$ (10$^{10}$) & 10 & 10 & 10 \\ 
f$_{c}$ \ \ \ (MHz) & 30 & 30 & 30 \\ 
$\beta _{e}/\beta _{p}$ \ \ \ \ (cm) & 0.25/0.25 & 0.25/0.25 & 0.25/0.25 \\ 
$\sigma _{e}/\sigma _{p}$ \ \ \ \ ($\mu m$) & 1.18/2.38 & 1.27/2.21 & 
1.38/2.02 \\ 
$\varepsilon _{e}^{N}/\varepsilon _{p}^{N}$ \ \ \ \ ($\mu mraad$)\  & 
1.1/15.8 & 1.5/11.5 & 2.15/8 \\ 
$\sigma _{ze}/\sigma _{zp}$ \ \ (cm) & 0.1/0.1 & 0.1/0.1 & 0.1/0.1 \\ 
$\Delta Q$ & 0.057 & 0.058 & 0.059 \\ 
D & 25.34 & 24.82 & 24.78 \\ 
H$_{D}$ & 1.24 & 1.38 & 1.49 \\ 
L \ \ \ (10$^{34}$ cm$^{-2}s^{-1}$) & 0.84 & 1 & 1.25 \\ 
Linac beam power \ \ \ (MW) & 4.8 & 5.66 & 6.8
\end{tabular}
\end{table}%
%

\begin{table}[tbp] \centering%
%
\caption{ Main parameters of $\tau $ factory\label{key}}

\bigskip 
\begin{tabular}{ll}
E$_{e}$ (GeV) & 2.1 \\ 
E$_{p}$ (GeV) & 2.1 \\ 
N$_{e}$ (10$^{10}$) & 0.07 \\ 
N$_{p}$ (10$^{10}$) & 10 \\ 
f$_{c}$ \ \ \ (MHz) & 30 \\ 
$\beta _{e}/\beta _{p}$ \ \ \ \ (cm) & 0.25/0.25 \\ 
$\sigma _{e}/\sigma _{p}$\ \ \ \ \ ($\mu m$) & 1.28/1.65 \\ 
$\varepsilon _{e}^{N}/\varepsilon _{p}^{N}$ \ \ \ \ ($\mu mraad$)\  & 2.7/4.5
\\ 
$\sigma _{ze}/\sigma _{zp}$ \ \ (cm) & 0.1/0.1 \\ 
$\Delta Q$ & 0.057 \\ 
D & 25 \\ 
H$_{D}$ & 1.69 \\ 
L \ \ \ (10$^{34}$ cm$^{-2}s^{-1}$) & 1.29 \\ 
Linac beam power \ \ \ (MW) & 7
\end{tabular}
\end{table}%
%

\bigskip

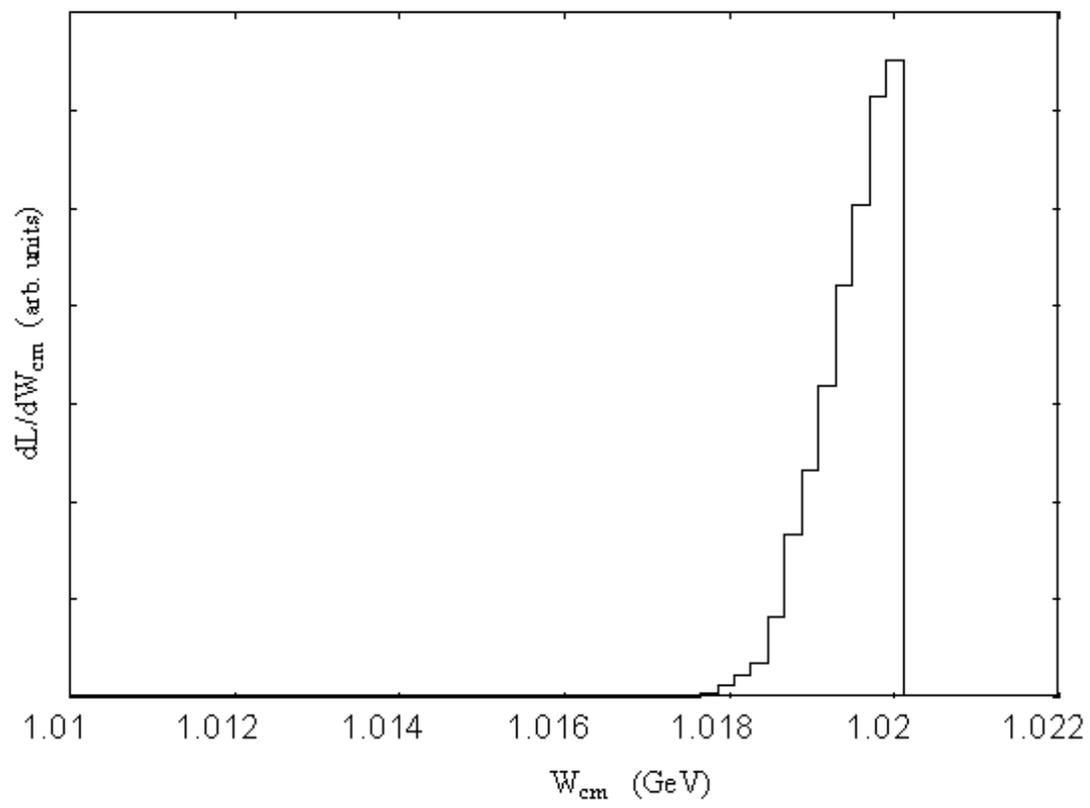
\begin{figure}[tbp] \centering%
%
\caption{Luminosity spectrum for $\phi $ factory (option C).\label{fig1}}%
\end{figure}%
%

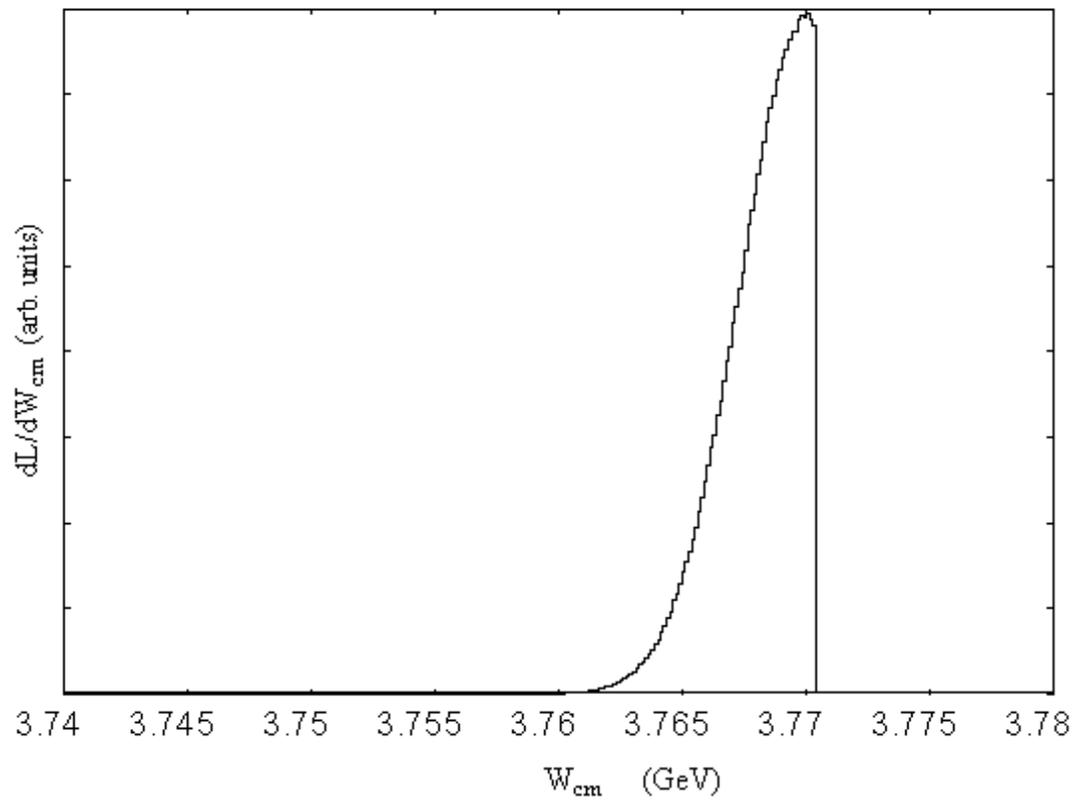
\begin{figure}[tbp] \centering%
%
\caption{Luminosity spectrum for charm factory (option C).\label{fig2}}%
\end{figure}%
%

\end{document}